# A NOVEL MULTIFACTOR AUTHENTICATION SYSTEM ENSURING USABILITY AND SECURITY

Gloriya Mathew*, Shiney Thomas**

*PG Scholar, **Assistant Professor
Department of Computer Science & Engineering
Amal Jyothi College of Engineerng, Kanjirappally
*gloriya007@gmail.com, shineythomas@amaljyothi.ac.in*

## ABSTRACT

*User authentication is one of the most important part of information security. Computer security most commonly depends on passwords to authenticate human users. Password authentication systems will be either been usable but not secure, or secure but not usable. While there are different types of authentication systems available alphanumeric password is the most commonly used authentication mechanism. But this method has significant drawbacks. An alternative solution to the text based authentication is Graphical User Authentication based on the fact that humans tends to remember images better than text. Graphical password authentication systems provide passwords which are easy to be created and remembered by the user. However, the main issues of simple graphical password techniques are shoulder surfing attack and image gallery attack. Studies reveals that most of the graphical passwords are either secure but not usable or usable but not secure. In this paper, a new technique that uses cued click point graphical password method along with the one-time session key is proposed. The goal is to propose a new authentication mechanism using graphical password to achieve higher security and better usability levels. The result of the system testing is evaluated and it reveals that the proposed system ensures security and usability to a great extent.*

## 1. INTRODUCTION

Authentication is the process to allow users to confirm his or her identity to a Web application. Human factors are considered to be the weakest part in a computer security system. The three major areas where human computer interaction is important are: authentication, security operations, and developing secure systems. Here we focus on the authentication problem. A password is a secret authentication data which is used to control access to a resource. The password is kept secret from those who are not allowed access, and those who wish to gain access to the resource are tested on whether or not they know the password and are granted or denied access accordingly.

Traditional textual password or PIN, however, relies on keyboard as the input device. Many researchers thereby look at an alternative approach graphical password. The password input is convenient as well as it is more user friendly in terms of memorability and recallability. The basic hypothesis is that human brain is more capable of storing graphical information than numbers or alphabets; in addition graphical password utilizes an easier and more human friendly memorization strategy recognition based memory, instead of recall based memory for textual password.

In this paper, a new authentication system which combines the advantages of both graphical password authentication system and one time session key is proposed. The system uses user defined images as image passwords and system defined pictures are used as decoy images. A random number key is generated by the system and using a GSM modem the key is send to the users mobile. This key is used only for one login session. The user needs to remember only the image passwords and its status uploaded by him during registration.

The rest of the paper is organized in the following way. In section 2, we provide a brief review of image based passwords and random number key generation. Then, the proposed system implementation is described in section 3. Section 4, describes the security analysis of the proposed system. Section 5 addresses the future work and concludes the paper.

## 2. RELATED WORKS

Authentication is a process which allows a user to confirm his identity to an application. It provides access control and user accountability. The problems with the text based passwords are well known. Users often create memorable passwords which are easy for attackers to guess, but strong system assigned passwords are difficult for the users to remember. An authentication system should encourage strong passwords while maintaining usability and memorability.

### 2.1 Authentication Methods

Authentication methods are broadly classified into three main areas. Token based authentication (two factor), Biometric based authentication (three factor), and Knowledge based authentication (single factor) [2][10].

#### 2.1.1 Token Based Authentication

Token based authentication is based on "Something You Possess". For example Smart Cards, a driver's license, credit card, a university ID card etc. In this method a token is used to access a specific resource by the user. Many token based authentication systems also use knowledge based authentication techniques to enhance security.

#### 2.1.2 Biometric Based Authentication

Biometric authentication system uses physiological or behavioural characteristics of a person for authentication. It is based on "Something You Are". Some of the biometric authentication systems use fingerprint recognition, face recognition, iris recognition or voice recognition to authenticate the users. Biometric identification depends on computer algorithms to make a yes or no decision. This method enhances user service by providing quick and easy identification.

#### 2.1.3 Knowledge Based Authentication

The knowledge based techniques are the most commonly used authentication systems. They are of two types: Text based passwords and picture based passwords. Although there are different types of authentication techniques available alphanumeric passwords are the widely used because they are versatile and it is easy to implement and use. The text based passwords need to satisfy two contradictory requirements. That is it should be easily remembered by the user and it should be hard to guess by an attacker. So these text passwords are vulnerable to dictionary attacks and brute force attacks. Knowledge based authentication can be used along with other authentication techniques to ensure security. Picture based passwords use images for creating passwords.

## 2.2 Image Based Password

Image based password (graphical password) refers to using pictures as passwords. It is an alternative to text based password. Graphical passwords are easier to remember because people remember pictures better than text. Visual objects offer a much larger set of usable passwords. Psychological studies reveals that people can remember pictures better than text; especially the photos, which are easier to remember than random pictures [2]. For example we can recognize the people we know easily from thousands of faces; this fact was used to implement a new authentication system. Since the search space of an image password is practically infinite it is resistant to brute force attack. An excellent survey of the numerous graphical password schemes has been developed [5]. In general, graphical passwords techniques are classified into three types:

- Recognition based techniques
- Recall based techniques
- Cued recall based techniques

In recognition-based techniques, a user is authenticated by challenging him/her to identify one or more images (e.g., faces, random art images, images clustered into semantic categories) he or she chooses during the registration stage. In a recall based technique a user is asked to repeat something that he is created or selected during the registration stage (e.g. draw a secret method).

## 2.3 Cued Recall Based Technique

In a cued recall based graphical password systems, users identify and target previously selected locations within one or more images. The images itself act as memory cues to aid recall [6].

Example systems:

- Passpoints

- Cued click points

### 2.3.1 Passpoints

In passpoint technique, password contains sequence of click points on a given image. The image is divided into tolerance squares. Users can select any points on the image as password in any order during registration stage[7][8]. To login to the system the users have to correctly click on the same points in the image in correct order which is in the same tolerance squares as entered in the registration stage. The main drawback of this technique is hotspot problem.

### 2.3.2 Cued Click Points

Cued Click points (CCP) is a technique which reduces the drawbacks of the passpoints. In this method users have to click on one point per image on five different images shown in sequence [9]. When we click on one point on an image another image is displayed. That is the next image displayed by the system depends on the previously selected click point. To create a different password users have to click on different click points in different images. This method provides implicit feedback which will be useful only to the legitimate users. While logging on users must have to click on the same click points in the sequence of image. If any of the click points selected by the user is not correct, authentication failure is indicated after selecting the final click point only [11].

## 2.4 Persuasive Technology

Persuasive technology is an approach to influence the people to behave in a desired manner. This technology is used to create stronger passwords. This method will encourage the users to select random points in an image for creating stronger password [4]. Persuasive technology reduces the hotspot problem in image passwords and reduces the drawbacks of other knowledge based authentication systems. Two types of persuasive technologies are: Persuasive text based password and Persuasive cued click point [1][3].

### 2.4.1 Persuasive Cued Click Points (PCCP)

User testing and analysis results showed that hotspots remained a problem with CCP. To overcome the hotspot problem persuasive feature is added to CCP. PCCP is obtained by adding a persuasive feature to CCP. PCCP [1] will encourage users to select less predictable passwords and thus reduces hotspot problem. A viewport is used in PCCP method. During the registration stage the image is shaded slightly except the viewport. The viewport is positioned randomly to avoid hotspots problem [11]. Users must select a click point within the viewport and cannot click outside of the viewport. A shuffle button is provided to randomly reposition the viewport. Shuffle button and viewport is existed only in the registration stage. During the login stage, the images are displayed normally. The users have to correctly click on the tolerance squares of the previously selected points on the image during registration stage.

The authentication using PCCP is highly secure but it is not that much usable. This will leads to develop the proposed system which ensures both the security as well as usability.

## 2.5 One Time Password

One time password (OTP) is a password which is used for a login time only. That means, it is valid only for one transaction. OTP can be used along with other techniques to enhance the security of the authentication systems. Replay attacks can be avoided using OTPs. In the proposed system a random number OTP is generated by the system during each login session and is send to the users mobile using a GSM modem.

## 3. PROPOSED SYSTEM

The proposed system is a combination of graphical password and a one-time session key (one-time password). The system allows the user to select three image passwords one in each level. When the user uploads the image password, the system will divide the uploaded image into a 3x3 grids. The system will provide four options (status) for labelling the grids. The four options are: Left to Right, Right to Left, Top to Bottom and Bottom to Top. The user can choose any of these options for each of the image password. The user must have to remember the uploaded image & its status that he chooses as the password during login stage. A random number session key (One Time Password-OTP) is generated by the system and send to users mobile for login to the system. The OTP consists of three random numbers which indicates the grid to be clicked on image password in each level of login stage. Proposed system architecture is shown in Figure 1. The proposed system consists of three modules:

- User Registration
- Image Selection
- Login

The proposed authentication system works as follows: User registration phase includes the registering the users to the system. For registering to the system user have to click on the "new user registration" button. It will display a registration form. For registration the users have to first enter his username and the other relevant data that is given in the registration form. The username must be unique to the system. The system will check whether the username is already existing in the database or not. The details of the user obtained from the registration form are stored in the database. After filling the registration form successfully, the user enters to the "image selection" module.

In the image selection phase, the user creates a graphical password by first uploading a picture he or she chooses from his own system using "UploadImage1" button. The user then chooses any one status from given four options: Left-right, Right-Left, Top-bottom, and Bottom-top. The system will then divide the selected picture into a 3x3 grid and label each grid according to the selected status. When the user click on the "Next" button the window for creating image password level-2 is displayed. In this window user have to click on to the "UploadImage2" button to select the second picture as the next image password. After selecting the picture the user must have to choose the option for labelling the grids in the picture. Then user click on to the "Next" button once more to select the third picture (image password) as in the previous levels. Finally click on the "Finish" button to complete the registration phase. Figure 2 shows a screenshot for creating image password in registration phase.

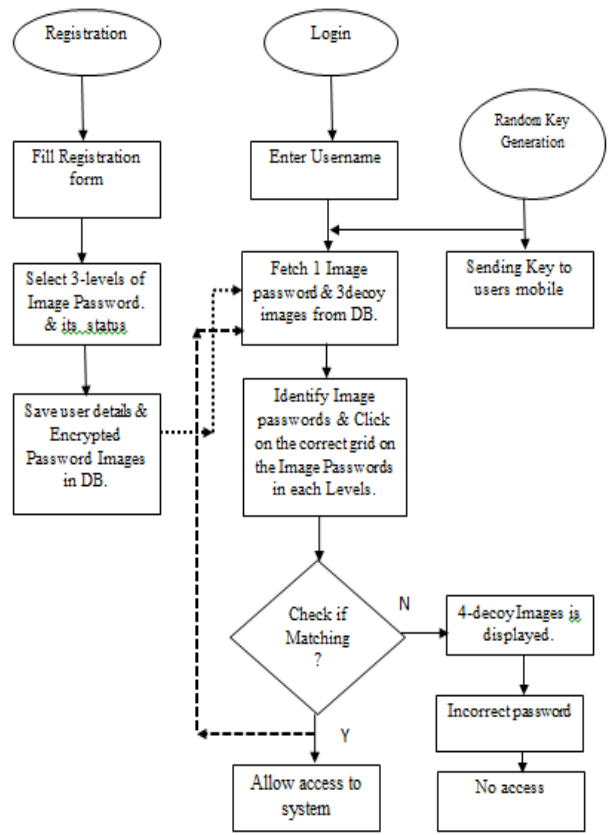

Figure 1: Proposed System Architecture

For authentication (Login) the user first enters his unique userid (username). Then click on the "Next" button. At the same time a one-time random number key is issued by the system and send to the user's mobile number given at the registration stage. For example suppose the key obtained to the users mobile is 386. Now the system displays four images which are not labelled. One among this is the first image uploaded by the user and rest of the three images are decoy images displayed by the system for confusioning the hacker. Since the key is 386, the user must have to click on the grid 3 on the actual picture (first image uploaded) among the four images. Then click on "Next" button. Now another set of four images is displayed. Among these four images one will be the second image uploaded by the user during the registration stage and the other three images will be displayed by the system for protecting from hacker. From these four images the users have to correctly click on the grid 8 on the second image uploaded by him. Similarly, when we click on "Next" button another set of four images is displayed. Among these four images one will be the third image uploaded by the user and three images will be displayed by the system. The user must have to correctly click on the grid 5 according to the grid labelling option (status) given to this image during the registration phase. If all the clicks in each level of images are correct then user can successfully logon to the system. Otherwise if there is any mistake in any of the click point (grid no.) system will displays an error message to the user. Figure 3, shows the screenshot of the login screen.

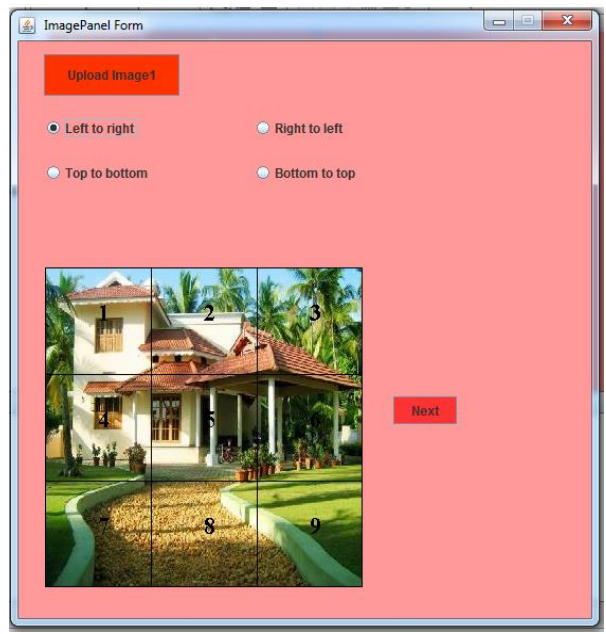

Figure2: Creating image password in registration phase.

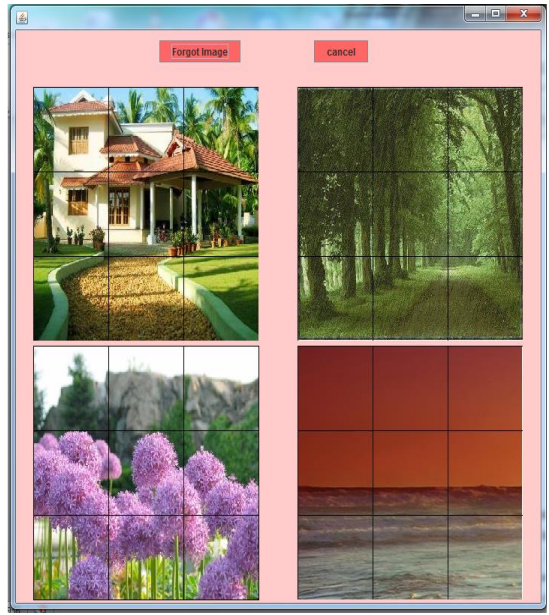

Figure 3: Login screen

## 4. SECURITY ANALYSIS

The inclusion of one time password along with cued click based graphical password method has improved the system performance to a great extent. It has an added advantage of the One-time session key and cued click point authentication systems. There is no possible method to break the system using cameras, key loggers and mouse detection software's. The proposed system will overcome hotspot problem seen in many of the graphical password authentication systems. The proposed system is secure because the password of the user does not known by anyone because the OTP is obtained to the users own mobile. The screen of mobile is small which decreases the visibility view of OTP to other users. This will decrease the shoulder surfing problem. Also, only the user knows the image password he uploaded and its status. All these factors will ensure the security of the system.

User Satisfaction was tested against the overall user interface of the system and whether or not the users are willing to use such an authentication system from then onwards or in the future. The user's opinion about this system is collected by using the questionnaires. By evaluating the response obtained for the questionnaires it reveals that more than 80 percentage of the users are satisfied with the new system.

In order to evaluate the usability and security of the system questionnaires were prepared and system testing is conducted using a group of 15 students in the age between 20 and 30. Table 1 indicates that the users take more time for registration since they were new to the system. The login time is reduced in each time when they login to the system. That is the login time for the first time (Login time-1) takes more time. Whereas the successive login to the system will reduce the login time of each user. This result reveals that the users feel convenient in using the system.

| Sl.No | Registration Time(s) | Login Time-1(s) | Login Time-2(s) | Login Time-3(s) |
|---|---|---|---|---|
| 1 | 76 | 55 | 37 | 29 |
| 2 | 104 | 36 | 17 | 15 |
| 3 | 86 | 31 | 17 | 13 |
| 4 | 102 | 49 | 34 | 21 |
| 5 | 166 | 51 | 22 | 16 |
| 6 | 88 | 18 | 25 | 12 |
| 7 | 43 | 13 | 14 | 12 |
| 8 | 91 | 43 | 17 | 14 |
| 9 | 113 | 20 | 17 | 12 |
| 10 | 70 | 26 | 15 | 14 |
| 11 | 40 | 16 | 13 | 8 |
| 12 | 119 | 34 | 31 | 17 |
| 13 | 62 | 23 | 15 | 10 |
| 14 | 54 | 31 | 19 | 12 |
| 15 | 68 | 42 | 38 | 19 |

Table 1: Registration time and login time taken by the users during system testing.

Questionnaires were used as a tool for the purpose of collecting the feedback from users during the system testing. The response to the questionnaires reveals that above 80 percentage of the users selected for the testing purpose is satisfied with the system. Some users feel some difficulty during the login time for remembering the status of image they selected. But, almost 80 percentage of the user's succeeded in login to the system. While evaluating the testing results it shows that system will provide enough security as well as usability.

## 4.1 Password Space

The theoretical password space of the proposed system can be expressed as shown below:
$(((w*h)/t^2)*m)*n)^c$ where $w*h$ is the size of the image password, $t^2$ is the size of the grid, $m$ is the number of images displayed in a level, $n$ is the number of statuses (in this system $m=n=4$) and $c$ is the number of click points (or levels) in the system (in this system $c=3$). Since, OTP is a random number it also helps to increase the password space of the proposed system. OTP is obtained in users mobile which have only a small screen to view the OTP, it will reduce the visibility view of the OTP and thus the proposed system reduces the chance of shoulder surfing problem to a great extent. The relationship between security and usability of the proposed system is shown in Figure 4.

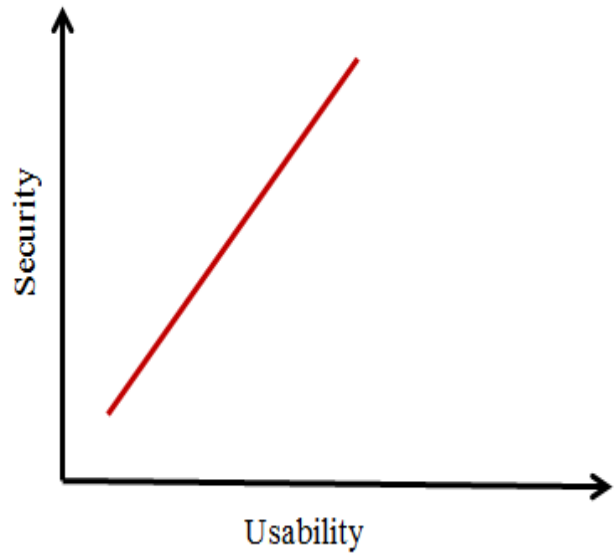

Figure 4: Relation between usability and security of the proposed system.

## 5. CONCLUSION

Graphical passwords are strong alternatives to text based and biometric authentications. Brute force search and dictionary attack is difficult in graphical passwords because search space is infinite. In this paper, a secure graphical password authentication system is proposed. The system combines graphical password and one-time session key trying to achieve the best of both methods which will increases the security. This system maintains some aspects of existing authentication mechanisms and will improve security and usability. Also the system overcomes the disadvantages of most of the graphical password techniques. Results show that this system provides a higher password space which ensures higher security and from the response obtained to questionnaires indicates that the system is also usable. The system has adopted the cued click point technique which offers attractive usability properties, such as cueing and good memorability. A random number one time password is generated in each login stage which cannot be used again will enhances the security of the system. The OTP used in the system will reduces the chance of shoulder surfing problem. Also the system stores the encrypted password images in the database which will prevent the image gallery attacks. The proposed system can be used in desktop locking applications, network security as well as for web security and other high security applications.